\documentclass[12pt]{article}

\input amssym.def
\input amssym.tex

\begin{document}

\begin{titlepage}

\begin{flushright}
\end{flushright}
\vskip 2.5cm

\begin{center}
{\Large \bf Coupling Right- and Left-Handed Photons\\
Differently to Charged Matter}
\end{center}

\vspace{1ex}

\begin{center}
{\large Brett Altschul\footnote{{\tt baltschu@physics.sc.edu}}}

\vspace{5mm}
{\sl Department of Physics and Astronomy} \\
{\sl University of South Carolina} \\
{\sl Columbia, SC 29208 USA} \\

\end{center}

\vspace{2.5ex}

\medskip

\centerline {\bf Abstract}

\bigskip

We consider a modification of electrodynamics in which right- and left-circularly
polarized photons are coupled to charged sources differently. Even though photon
helicity is a Lorentz invariant quantity, such a modification breaks Lorentz
symmetry, as well as locality. The modified theory includes novel magnetic forces
between perpendicular currents. Existing data can be used to constrain the
modification at approximately a $2\times10^{-3}$ level.

\bigskip

\end{titlepage}

\newpage

\section{Introduction}

In recent years, there has been a great surge in interest in the possibility of
Lorentz
violation~\cite{ref-reviews}. The exploration of possible deviations from special
relativity has both important experimental and theoretical facets. There are many
different forms that Lorentz violation could take and several different
approaches to the questions surrounding them.

There are a number of possible motivations for focusing on a particular
modification of known physics.
The minimal SME approach~\cite{ref-kost2} considers a theory containing all possible
local,
gauge-invariant, superficially renormalizable operators involving only standard model
fields---even when the operators are not Lorentz invariant.
The number of such operators, unrestricted by Lorentz symmetry, is quite
large. Alternatively, one might consider, in a
unsystematic way, only certain operators that have particularly interesting or
attractive structures.

In this paper, we shall look at a form of Lorentz violation that has been selected
for yet a different reason. Rather than beginning with a particular form of
interaction, we start with a physical phenomenon and ask what kind of interaction
could produce it. There are immediately two lines of inquiry that can be followed:
examining what impact the existence of the novel phenomenon must have in other
regimes and determining what are the experimental limits on the phenomenon. The first
line of inquiry entails developing a predictive quantum theory that contains the
novel phenomenon. In general, the choice of such a theory will not be unique.
However, there may exist a particular version which is clearly the simplest or has
the most desirable properties, whereas other formulations may turn out to have pathologies.

In 1990, Carroll, Field, and Jackiw introduced a Lorentz-violating modification of
electromagnetism with the gauge field Lagrange density
\begin{equation}
{\cal L}_{CS}=-\frac{1}{4}F^{\mu\nu}F_{\mu\nu}+\frac{1}{2}k_{\mu}\epsilon^{\mu\nu
\rho\sigma}F_{\nu\rho}A_{\sigma}.
\end{equation}
$k$ is an externally prescribed four-vector.
This modification, which has a Chern-Simons
form~\cite{ref-carroll1,ref-jackiw2,ref-schonfeld}, causes right- and left-circularly
polarized electromagnetic waves to propagate at different phase speeds. The
birefringence that would result from such an effect has not been seen, even for
light coming from cosmological
distances~\cite{ref-carroll1,ref-carroll2,ref-goldhaber}.
Yet while this kind of Chern-Simons term is not experimentally viable, it has
stirred a great deal of interest in similar theories.

Searching for an analogous modification of general relativity, Jackiw and
Pi~\cite{ref-jackiw6} looked
at a model (which was also considered in \cite{ref-lue})
with a gravitational Lagrange density
\begin{equation}
{\cal L}_{G}=\frac{1}{16\pi G}\left[\sqrt{-g}R-\frac{1}{2}v_{\mu}\epsilon^{\mu\nu
\rho\sigma}\left(\Gamma^{\alpha}_{\nu\beta}\partial_{\rho}\Gamma^{\beta}_{\sigma
\alpha}+\frac{2}{3}\Gamma^{\alpha}_{\nu\beta}\Gamma^{\beta}_{\rho\gamma}\Gamma
^{\gamma}_{\sigma\alpha}\right)\right],
\end{equation}
which contains the analogous gravitational Chern-Simons term. (The divergence of the
quantity contracted with $v$ is proportional to the Riemann tensor contracted with
its dual.) However, in this case, if diffeomorphism invariance is enforced, the
apparent Lorentz violation coming from the prescribed $v$ is
illusory~\cite{ref-guarrera}; Lorentz
symmetry cannot be broken in a diffeomorphism invariant theory~\cite{ref-kost12}.
Nonetheless, the added term does have physical consequences. The two polarizations
of gravitational waves must travel at the same speed because of boost invariance,
but the circularly polarized radiation
states couple differently to the energy-momentum
tensor. In accordance with the appearance of the Levi-Civita tensor $\epsilon$ in
the modified action, this modification violates parity.

So, inspired by the electromagnetic Chern-Simons theory (in which the two
polarizations propagate differently) a similar-looking gravitational theory (in which
the polarizations couple differently) was introduced. The goal of this paper is to
examine a somewhat analogous theory to the one described by ${\cal L}_{G}$, but back
in the realm of quantum electrodynamics (QED). That is, we shall look at a
modification of QED that causes the two photon helicities to couple differently to
charged matter. Taking the simplest such model, we shall find that our model, like the
Chern-Simons theory ${\cal L}_{CS}$, violates Lorentz boost symmetry.

This paper is organized as follows. In section~\ref{sec-theory}, the basic theory
with different couplings for electromagnetic waves of different helicities is discussed. Section~\ref{sec-prop} shows how the novel interaction may be recast as a
change to the photon propagator. Then in section~\ref{sec-potential}, the modified
propagator is used to derive an expression for a new potential existing
between perpendicular current elements. Section~\ref{sec-constr} discusses the
question of experimental constraints on the modified theory and presents the paper's
final conclusions.

\section{QED with Helicity-Dependent Couplings}
\label{sec-theory}

In the Chern-Simons gravity theory described by ${\cal L}_{G}$, the coupling of the
gravitons to their sources depends not only on helicity but also on frequency. This
introduces a new dimensional scale $\mu$. Since gravitation is already described by
a nonenormalizable theory, the introduction of such a scale, accompanied in the
action by additional derivatives, does not necessarily
worsen the behavior of the theory.
However, QED is a renormalizable theory, and adding extra
derivatives to the coupling term would presumably destroy this feature; the
introduction an operator with mass dimension greater than four would generate
pathologies. For this
reason, we shall consider a modification of electromagnetism in which the photon
coupling depends on helicity but not on frequency and in which there is no new
dimensional scale.
(A similar modification of gravity was considered in~\cite{ref-contaldi}. The theory
had the differing right- and left-handed couplings of the Chern-Simons theory, but
without a new dimensional scale. Such a modification to gravity could lead to changes
to the cosmic microwave background polarization.)

The division of photons into right- and left-circularly polarized is superficially
Lorentz invariant. No rotation or boost will change the helicity of a given photon
mode. Nonetheless,
it does not appear to be possible to assign different couplings to the right- and
left-handed photons in Lorentz-invariant fashion. The reason is that the photon
portion of the electromagnetic field cannot be disentangled from the
electrostatic portion,
and there is no analogous Lorentz-invariant separation of the electrostatic
interaction. It is possible to separate the full electromagnetic sector into right-
and left-handed parts in the absence of charges, when the
electrostatic part of the Hamiltonian vanishes. However,
what remains is a free theory, for which the coupling is not an
observable parameter.

Our starting point will be the electromagnetic Hamiltonian in the Coulomb gauge,
$\vec{\nabla}\cdot\vec{A}=0$, which is
\begin{eqnarray}
H & = & H_{0}+H_{1}+H_{2} \\
 & = & \frac{1}{2}\int d^{3}x\left(\left|\vec{\nabla}\times\vec{A}\right|^{2}+\left|
\partial\vec{A}/\partial t\right|^{2}\right)-
\int d^{3}x\,\vec{\jmath}\cdot\vec{A}+
\frac{1}{2}\int d^{3}x\int d^{3}x'\frac{\rho(\vec{x},t)\rho(\vec{x}\,',t)}
{4\pi\left|\vec{x}-\vec{x}\,'\right|}. \nonumber
\end{eqnarray}
When the charge density is carried by
elementary charged particles, $H_{1}$ is ${\cal O}(e)$ while $H_{2}$ is already
${\cal O}(e^{2})$. We use this particular form of $H$ because it separates the
electrostatic and propagating parts of the electromagnetic field.  Such a separation
is needed, because the electrostatic field does not have a decomposition into
right- and left-handed parts the way the radiation field does.

In fact, in this gauge the electrostatic potential is a constrained degree of
freedom.
$A_{0}$ has been eliminated from the Hamiltonian in favor of the double integral
term $H_{2}$. (Note that
for point charges, this Coulomb repulsion term includes the infinite self-energies of
individual particles.) The remaining
$\vec{A}$ is constrained by the Coulomb condition, leaving two physical degrees of
freedom. These two degrees of freedom contain the photon modes as well as all
magnetostatic effects.
The residual gauge symmetry of this Hamiltonian is that we may add to $\vec{A}$ the
gradient of a harmonic function.

Our modification shall be splitting the photon field into two separate parts, with
different helicities, and coupling them differently to the current $\vec{\jmath}$.
Taking
\begin{equation}
\vec{A}(\vec{x},t)=\vec{A}_{+}(\vec{x},t)+\vec{A}_{-}(\vec{x},t),
\end{equation}
the two halves of $\vec{A}$ are
\begin{equation}
\vec{A}_{\pm}(\vec{x},t)=\int\frac{d^{3}q}{(2\pi)^{3}}\frac{1}{2\omega_{\vec{q}}}
\left[a_{\vec{q},\pm}\vec{\epsilon}\,^{(\pm)}e^{-iq\cdot x}+
a^{\dag}_{\vec{q},\pm}\vec{\epsilon}\,^{(\pm) *}e^{iq\cdot x}\right].
\end{equation}
The circular polarization vectors are $\vec{\epsilon}\,^{(\pm)}(\vec{q}\,)=\mp\frac
{1}{\sqrt{2}}\left[\vec{\epsilon}\,^{(1)}\pm i\vec{\epsilon}\,^{(2)}\right]$, where
$\left[\vec{\epsilon}\,^{(1)},\vec{\epsilon}\,^{(2)},\hat{q}\right]$ form a
right-handed triplet; and $q_{0}$ in the four-vector dot product $q\cdot x$ is
$q_{0}=\omega_{\vec{q}}=|\vec{q}\,|$.

We then replace the conventional
$H_{1}$ with
\begin{equation}
\label{eq-H1prime}
H_{1}'=-\sum_{\pm}\frac{(1\pm\kappa)}{\sqrt{1+\kappa^{2}}}
\int d^{3}x\,\vec{\jmath}\cdot\vec{A}_{\pm}.
\end{equation}
Thus right- and left-handed photons couple to the current with different strengths;
the difference is determined by the (small) parameter $\kappa$. The insertion of
$1/\sqrt{1+\kappa^{2}}$ might seem peculiar, but it is related to the fact that
there are actually three coupling constants present in the the theory now---the
couplings for right- and left-handed photons, $e_{(+)}$ and $e_{(-)}$, and also the
electrostatic coupling $e_{(0)}$ that appears in $H_{2}$. It seems obvious that
$e_{(0)}$ should be some kind of average of $e_{(+)}$ and $e_{(-)}$, and as we shall
see, the most natural relationship among the three is
\begin{equation}
\label{eq-e0}
e_{(0)}^{2}=\frac{1}{2}[e_{(+)}^{2}+e_{(-)}^{2}].
\end{equation}
This relationship is embodied in the modified coupling (\ref{eq-H1prime}).

In studies of Lorentz violation, it is important to distinguish distinguish between two
types of Lorentz transformations. ``Observer'' (or ``passive'') transformations correspond
merely to relabeling of coordinates, and all theories should be invariant under this kind of
reparameterization. ``Particle'' (or ``active'') transformations, on the other hand, are physically
meaningful. While an observer rotation merely means studying an experiment in
rotated coordinates, the particle rotation corresponds to actually rotating the experimental
apparatus into a different orientation. The simplest phenomena in Lorentz-violating
physics are preferred frame effects. The preferred frame is one in which the physics are
invariant under rotations but not under boosts. The form of Lorentz violation discussed
here is clearly of the preferred frame type. More generally, if a type of Lorentz violation
may be completely parameterized by a single timelike four-vector $w^{\mu}$,
there is always an
isotropic preferred frame---the one in which the spatial components of $w$ vanish.

Moreover, the discrete symmetries of the modification are easy to determine. The
conventional electromagnetic coupling is invariant under parity (P), charge conjugation
(C), and time reversal (T). The new term added to the Hamiltonian has the same form,
except it includes an extra factor of the photon helicity. The helicity is odd under P and
even under C and T; the novel interaction inherits these same discrete symmetries and is
hence odd under CPT. This is actually not surprising, since we shall see that this form of
Lorentz violation can indeed be described by a single preferred timelike vector. The usual
expectation is that forms of Lorentz violation described by background tensors with odd
numbers of Lorentz indices should be odd under CPT, while those with even numbers of
indices should be even under CPT. While this correspondence does not necessarily hold
outside the scope of local Lagrangian field theory~\cite{ref-altschul14}, it does hold in this case.

\section{Electromagnetic Propagator}
\label{sec-prop}

In perturbation theory, the fundamental objects of interest are the propagators and
the vertex factors. The form of Lorentz violation considered here---although it
appears as a modification of the interaction---can be absorbed into a modified photon
propagator. In the
noncovariant approach based on the Hamiltonian $H$, the photon propagator must be
assembled from matrix elements of both $H_{1}$ and $H_{2}$. At ${\cal O}(e^{2})$,
a matrix element receives a second-order contribution from two factors of
$H_{1}$ and a first-order
contribution from $H_{2}$. These fit together to yield a covariant result.
(The procedure is outlined in~\cite{ref-sakurai}, for example.) The $H_{1}$
part of the calculation involves evaluating
\begin{equation}
\tilde{D}_{jk}(x-x')\equiv\langle 0|T\{A_{j}(x)A_{k}(x')\}|0\rangle=\int\frac{d^{4}q}
{(2\pi)^{4}}
\delta^{T}_{jk}\frac{i}{q^{2}+i\epsilon}e^{-iq\cdot(x-x')}.
\end{equation}
$\delta_{jk}^{T}$ is the transverse $\delta$-function, $\delta_{jk}^{T}=
\delta_{jk}-\hat{q}_{j}\hat{q}_{k}$.

With the parity-violating $H'=H_{0}+H_{1}'+H_{2}$, the only change to the
propagator calculation is that we
must evaluate
\begin{equation}
\tilde{D}_{jk}'(x-x')
=\langle 0|T\left\{\frac{(1+\kappa)^{2}}{1+\kappa^{2}}[A_{+}(x)]_{j}
[A_{+}(x')]_{k}+\frac{(1-\kappa)^{2}}{1+\kappa^{2}}[A_{-}(x)]_{j}[A(x')_{-}]_{k}
\right\}|0\rangle.
\end{equation}
(The cross terms with $A_{+}A_{-}$ vanish.)
In including the $\kappa$-dependent factors as part of
$\tilde{D}_{jk}'$, we are effectively absorbing these factors into the definition
of the electromagnetic field. This can
also be accomplished directly with a nonlocal field
redefinition, which moves the new physics from the interaction term $H_{1}$ to the
photon kinetic term $H_{0}$.

In passing from $\tilde{D}_{jk}$ to $\tilde{D}_{jk}'$, the
only modifications needed are to the expression's
Lorentz structure. For $\tilde{D}_{jk}'$,
we must evaluate
\begin{equation}
\label{eq-polsum}
\sum_{\pm}\frac{(1\pm\kappa)^{2}}{1+\kappa^{2}}\epsilon^{(\pm)}_{j}
\epsilon^{(\pm)*}_{k}=\delta_{jk}^{T}-2i\frac{\kappa}{1+\kappa^{2}}\epsilon_{jkl}
\hat{q}_{l}.
\end{equation}
The $\delta_{jk}^{T}$ term combines with the electrostatic contribution to give
the usual photon propagator; the $1/\sqrt{1+\kappa^{2}}$ in (\ref{eq-H1prime})
ensures the normalization of this contribution
is correct. The other term in (\ref{eq-polsum}) generates a
new contribution to the effective propagator.
In momentum space, this propagator is
\begin{equation}
\label{eq-Dprime}
D_{\mu\nu}'(q)=-i\frac{g_{\mu\nu}-
(1-\zeta)q_{\mu}q_{\nu}/q^{2}-2i\frac{\kappa}{1+\kappa^{2}}
\hat{q}^{\beta}\epsilon_{0\beta\mu\nu}}{q^{2}+i\epsilon}.
\end{equation}
The four-vector $\hat{q}^{\mu}$ can be either $(0,\hat{q})$ or $(q_{0}/|\vec{q}\,|,
\hat{q})=q^{\mu}/
|\vec{q}\,|$. The gauge parameter $\zeta$ appears here even though we derived the
propagator from a gauge-fixed Lagrangian, because the term it multiplies has no
effect when contracted with conserved currents.
If this were to be a viable modification of QED, the effective propagator
(\ref{eq-Dprime}) ought to describe essentially all the new physics that arise in
the model. The different couplings at
the vertices for the two helicities have been converted into differing field strengths
during propagation.

While (\ref{eq-Dprime}) provides a complete description of the tree-level photon
propagator, it is
also natural to ask how the electromagnetic two-point function is affected by quantum
corrections. At one-loop order (and leading order in $\kappa$), the photon self-energy diagram
contains no internal photon propagators, and so the vacuum polarization tensor
$\Pi^{\mu\nu}(q)=(q^{2}g^{\mu\nu}-q^{\mu}q^{\nu})\Pi(q^{2})$
is unchanged from the standard theory. In the Landau gauge
($\zeta=1$),
the insertion of a virtual fermion-antifermion loop into the photon propagator merely
multiplies the propagation amplitude
(\ref{eq-Dprime}) by the $q^{2}$-dependent renormalization factor
$\Pi(q^{2})$.
As a consequence, one-loop quantum corrections do not introduce any new Lorentz
structures in the two-point correlation function, nor do they affect the relative magnitudes of the Maxwell term and the novel $\kappa$-dependent term.


We note, moreover, that the new term in the
numerator of (\ref{eq-Dprime}) does not violate
gauge invariance in any obvious way. It is transverse, since $q^{\mu}
(\hat{q}^{\beta}\epsilon_{0\beta\mu\nu})=0$.
The new interactions were introduced into a Hamiltonian that was already gauge fixed,
so it is not possible to answer unambiguously whether they are gauge invariant.
However, the theory appear to have two physical photon polarizations for each
$\vec{q}$, a transverse propagator, and no new dimensional constants---all
characteristic of a renormalizable theory.

However, the theory definitely does not possess Lorentz invariance;
this is evident from the appearance of the specific index 0 in
$\epsilon_{0\beta\mu\nu}$. The propagator could be rewritten using an externally
prescribed four-vector $\kappa^{\mu}=\left(\kappa,\vec{0}\,\right)$, but the correct
generalization for a $\kappa^{\mu}$ that is spacelike is not obvious, because of the
presence of the unit three-vector $\hat{q}$.
However, for any timelike $\kappa^{\mu}$, it is possible to
perform an observer boost to eliminate the spatial components of $\kappa^{\mu}$,
leaving a theory with photon propagator $D_{\mu\nu}'$ in the boosted frame.

At leading order, this modification to the propagator is similar to a theory with
a term proportional to  $\epsilon_{0\beta\mu\nu}
(\partial^{\mu}A^{\delta})[(\partial^{\beta}/|\vec{\nabla}|)\partial^{\nu}
A_{\delta}]$ (which is weakly nonlocal)
added to the Lagrange density. There are
differences apparent at higher orders, however. This raises the question of
whether the theory discussed here can be considered local, and the answer is
slightly
ambiguous. The Hamiltonian $H$ is nonlocal in its electrostatic part, but this is
not the portion of the Hamiltonian that has been modified. Indeed, the contributions
made by
$H_{2}$ in perturbation theory contribute to the photon propagator in
exactly the same
way they do in conventional electrodynamics (which is certainly a local theory). The
changes made to $H_{1}'$ can be described in a completely local formalism;
for although
separating the vector potential $\vec{A}$ into its right- and left-circularly
polarized parts requires an expansion in Fourier modes, there is no reason we cannot
treat $\vec{A}_{+}$ and $\vec{A}_{-}$ as the fundamental fields, which interact
with $\vec{\jmath}$ in a completely local fashion. However, if these are the
fundamental fields, the electrostatic $H_{2}$ may no longer be viewed as a
manifestation of the same electromagnetic field as appears in the other terms in the
Hamiltonian. It would have to represent a new interaction, fundamentally nonlocal
in nature. Thus it does not appear possible to escape the nonlocality completely.

It seems that
locality and Lorentz symmetry are violated in similar ways. The gauge-fixed
Hamiltonian
$H$ has neither property, although it represents a theory that is ultimately
both local and Lorentz invariant. Superficially, $H_{1}-H_{1}'$, is both local and
Lorentz invariant, but introducing it interferes with the subtle interplay among
$H_{0}$, $H_{1}$, and $H_{2}$ that ensures a local, Lorentz-invariant $S$-matrix.

In fact, even Lorentz-violating field theories that are completely local can 
have problems with stability, causality, or both~\cite{ref-kost3}. This has
been worked out quite explicitly for the Chern-Simons theory described by
${\cal L}_{CS}$~\cite{ref-carroll1} with a timelike $k$. There are runaway solutions
to the equations
of motion, because the Hamiltonian is not bounded from below. One can arrange the
boundary conditions such that the runaway modes are never excited. However,
the Green's functions with these boundary conditions are acausal; charges begin to
radiate before they actually accelerate.

On the other hand, there are
nonlocal Lorentz-violating field theories that are better behaved with respect to
stability and causality than the local theories~\cite{ref-altschul14}. It is
difficult to see how the present theory could have problems with causality, since
the photon modes propagate only on the light cone. Stability of the theory as a
whole (including charged fermions) is not so clear, but there are certainly
no runaway modes in the pure electromagnetic sector.

\section{Anomalous Potential}
\label{sec-potential}

At nonrelativistic energies, the dominant effect of the electromagnetic field is the
Coulomb
interaction between charges; there is also a magnetostatic potential between idealized
infinitesimal current elements. Changes to the structure of the electromagnetic sector
will generally produce corresponding changes in the nonrelativistic potentials.
However,
in the modified theory discussed here, there are not expected to be any changes to the
scalar potential $A_{0}$; the Coulomb part of the Hamiltonian $H_{2}$
was explicit in the original formulation of the theory, and it was not modified. In
contrast, there is a change to the potential between currents, which we can evaluate.

We consider two infinitesimal current elements, $d\vec{I}_{1}=I_{1}\,d\vec{l}_{1}$
and $d\vec{I}_{2}=I_{2}\,d\vec{l}_{2}$,
separated by a vector $\vec{r}_{12}$. The nonrelativistic potential between them is the
three-dimensional Fourier transform of the contraction of these currents with the
effective propagator,
\begin{equation}
V(\vec{r}_{12})=i\int\frac{d^{3}q}{(2\pi)^{3}}e^{i\vec{q}\cdot\vec{r}_{12}}
\left[dI_{1}^{j}D'_{jk}(q_{0}=0,\vec{q}\,)dI_{2}^{k}\right].
\end{equation}
The $-ig_{\mu\nu}/(q^{2}+i\epsilon)$ term in the propagator gives rise to the usual
(doubly differential) potential between the current elements
$V_{0}(\vec{r}_{12})=-\frac{1}{4\pi r_{12}}(d\vec{I}_{1}\cdot d\vec{I}_{2})$. The
calculation follows precisely the same path as the calculation of the Coulomb
potential between pointlike charges. However, we are concerned with the novel term,
\begin{equation}
V_{\kappa}=2i\frac{\kappa}{1+\kappa^{2}}\int\frac{d^{3}q}{(2\pi)^{3}}
e^{i\vec{q}\cdot\vec{r}_{12}}
\frac{1}{\vec{q}\,^{2}-i\epsilon}[(d\vec{I}_{1}\times d\vec{I}_{2})
\cdot\hat{q}].
\end{equation}

Splitting $d\vec{I}_{1}\times d\vec{I}_{2}$ into its components parallel and normal
to $\vec{r}_{12}$, only $(d\vec{I}_{1}\times d\vec{I}_{2})_{\parallel}=
[(d\vec{I}_{1}\times d\vec{I}_{2})\cdot\hat{r}_{12}]\hat{r}_{12}$ will contribute.
This is evident from symmetry considerations alone;
the triple product $(d\vec{I}_{1}\times d\vec{I}_{2})\cdot\vec{r}_{12}$ is the only
pseudoscalar that can be constructed from $d\vec{I}_{1}\times d\vec{I}_{2}$.
Moreover, it is easy to see explicitly that for every direction $\hat{q}$, there is
another
one $\hat{q}'$, such that $\vec{q}\cdot\vec{r}_{12}=\vec{q}\,'\cdot\vec{r}_{12}$,
yet $(d\vec{I}_{1}\times d\vec{I}_{2})_{\perp}\cdot\hat{q}=
-(d\vec{I}_{1}\times d\vec{I}_{2})_{\perp}\cdot\hat{q}'$, where
$(d\vec{I}_{1}\times d\vec{I}_{2})_{\perp}=d\vec{I}_{1}\times d\vec{I}_{2}-
(d\vec{I}_{1}\times d\vec{I}_{2})_{\parallel}$; and
this leads to complete cancellation
in the $(d\vec{I}_{1}\times d\vec{I}_{2})_{\perp}$ part of the integral.

As in the evaluation of the nonrelativistic Coulomb potential,
we evaluate the integral in spherical coordinates and use $\hat{r}_{12}\cdot
\hat{q}=\cos\theta$. This gives
\begin{equation}
V_{\kappa}(\vec{r}_{12})=\frac{i}{2\pi^{2}}\frac{\kappa}{1+\kappa^{2}}
(d\vec{I}_{1}\times d\vec{I}_{2})\cdot\hat{r}_{12}\int_{0}^{\infty}dQ
\frac{Q^{2}}{Q^{2}-i\epsilon}\int_{-1}^{1}d(\cos\theta)\,
\cos\theta e^{iQr_{12}\cos\theta},
\end{equation}
where $Q$ denotes $|\vec{q}\,|$. The $i\epsilon$ prescription in the denominator is
unneeded, and performing the angular integration, we have
\begin{eqnarray}
V_{\kappa}(\vec{r}_{12}) & = & \frac{1}{2\pi^{2}}\frac{\kappa}{1+\kappa^{2}}
(d\vec{I}_{1}\times d\vec{I}_{2})\cdot\hat{r}_{12}
\left[\frac{2}{r_{12}^{2}}\int_{0}^{\infty}dQ\frac{\sin(Qr_{12})-Qr_{12}\cos
(Qr_{12})}{Q^{2}}\right] \\
& = & \frac{1}{\pi^{2}}\frac{\kappa}{1+\kappa^{2}}\frac{1}{r_{12}}
(d\vec{I}_{1}\times d\vec{I}_{2})\cdot\hat{r}_{12}.
\end{eqnarray}

Evidently, by endowing right- and left-circularly polarized photons with different
couplings to charged matter, we have also introduced a new potential between
perpendicular
currents. The structure of $V_{\kappa}$ clearly indicates that it is capable of
generating physical forces. For example, two current-carrying wires running in
perpendicular directions but not intersecting will feel an attractive or repulsive
force, depending on the sign of $(\vec{I}_{1}\times \vec{I}_{2})\cdot\hat{r}$,
where $\hat{r}$ is directed between the wires' points of closest approach.

In fact, for long wires of length $L$, oriented so that current $I_{1}$
flows in the $x$-direction and $I_{2}$ flows in the $y$-direction, with a closest
approach separation $\vec{r}=(0,0,z)$ between their midpoints,
the force is
\begin{eqnarray}
F & = & -\frac{d}{dz}\left[\frac{1}{\pi^{2}}\frac{\kappa}{1+\kappa^{2}}\int_{-L/2}^{L/2}
dx_{1}\int_{-L/2}^{L/2}dy_{2}\frac{I_{1}I_{2}z}{x_{1}^{2}+y_{2}^{2}+z^{2}}\right] \\
& \approx &-\frac{d}{dz}\left[\frac{1}{\pi}\kappa I_{1}I_{2}z\int_{-L/2}^{L/2}dx_{1}
\frac{1}{\sqrt{x_{1}^{2}+z^{2}}}\right] \\
\label{eq-force}
& \approx & -\frac{2}{\pi}\kappa I_{1}I_{2}\log\frac{L}{z},
\end{eqnarray}
for $0<z\ll L$.

The detailed structure of the potential $V_{\kappa}$ at ${\cal O}(\kappa^{2})$
depends on the particular relation (\ref{eq-e0}).
However, the behavior at leading order in $\kappa$ is free of any
ambiguities associated with the choice of normalization.
If the right- and left-handed couplings were normalized differently,
then in addition to
the ${\cal O}(\kappa)$ contribution to $V_{\kappa}$, there would also be changes at
${\cal O}(\kappa^{2})$ to the usual potential between parallel current elements.

\section{Experimental Constraints}
\label{sec-constr}

It should be possible to test for the presence of $\kappa$ in several different ways.
Since $\kappa$ is known to be small, it is reasonable to
work to leading order in $\kappa$ in
examining these tests. One test is quite obvious. Since the two helicities of
light couple differently to moving charges, there would be a systematic ellipticity
in the radiation from what would be expected to be a linearly polarized source. Such
a source would emit radiation with a true elliptiticy of $|\kappa|$. This would be
detected as an apparent ellipticity of $2|\kappa|$, because the polarization that is
produced more weakly is also more weakly coupled to the detector. (If a field
redefinition is used to move the Lorentz violation into the Maxwell propagation
Lagrangian, then $2|\kappa|$ becomes the true ellipticity. The observable effect
is the same in either case.)

Synchrotron radiation is frequently used for calibrating x-ray polarimeters.
The Compton polarimeter described in~\cite{ref-they} is not sensitive to the relative
phases of the perpendicular polarization components it measures, so it interprets
elliptically polarized radiation as having a fictitious linear polarization.
The plane of this fictitious linear polarization can be identified to within
$4\times 10^{-3}$ radians accuracy (assuming conventional electrodynamics), and the
measurements made with this device are in agreement with the standard predictions at
this level. This sets a limit of $|\kappa|\lesssim 2\times 10^{-3}$.

More accurate
polarimetric measurements have been made in experiments looking at photon
birefringence (ref. \cite{ref-sushkov} measured ellipticities  at the $10^{-7}$
radians
level), but these are typically not sensitive to $\kappa$. There are systematic
effects (such as stress-induced birefringence in the
optical windows that open onto sample cells) with the same experimental signatures
as $\kappa$ that must be subtracted away; such subtraction obviously eliminates any
sensitivity to $\kappa$. However, the kinds
of apparatus used to make these birefringence measurements might be adapted to look
for a nonzero $\kappa$.

There are also potential magnetostatic tests. By running
two perpendicular current-carrying wires close together and measuring the forces they
exert on one-another, it would be possible to test for the presence of the
force (\ref{eq-force}). Searches for this kind of novel force could perhaps be
combined with searches for
other manifestations of Lorentz violation in electromagenticstatics, which can
involve preferred direction effects and mixing between electric and magnetic sources
and fields~\cite{ref-kost17}.

For comparison, the best laboratory bound on the Lorentz-violating coefficient
$\tilde{\kappa}_{{\rm  tr}}$---which is another isotropic boost invariance violation
parameter that may be introduced into the photon sector---are at the
$5\times10^{-15}$ level~\cite{ref-altschul20}. In the presence of
$\tilde{\kappa}_{{\rm  tr}}$, the speed of light becomes
$1-\tilde{\kappa}_{{\rm  tr}}$, and this affects the rate of synchrotron emission by
charged particles.
The $5\times10^{-15}$ bound derives from an analysis of synchrotron losses in energy calibration
data from the Large Electron-Positron Collider (LEP). The difference in sensitivities
to $\kappa$ and $\tilde{\kappa}_{{\rm  tr}}$
has a straightforward explanation. Any experimental constraint on $\kappa$ requires
a measurement of a P-odd observable. Consequently, many types of experiments are
insensitive to $\kappa$. For example, the energy loss by a lepton beam during a full
revolution around the LEP ring is unaffected by $\kappa$.

The current-current interactions deriving from $\kappa$ also have minimal effects on
the mutual interactions of magnetic moments. The field of a circulating current is
well described by a magnetic dipole at distances large compared to the size of the
current loop. At these distances, the currents on opposite sides of the loop make
almost equal and opposite contributions to any force exerted on the dipole due by the
new current-current
interaction. Conversely, a dipolar field will exert minimal novel forces on other
currents in the vicinity; for idealized
pointlike dipoles, the $\kappa$-dependent forces cancel completely. This makes it
difficult to constrain $\kappa$ with precision atomic experiments, which frequently
measure the interactions between dipoles, and this insensitivity is a further
consequence of the P-odd character of the $\kappa$ interaction.

Direct measurements of the effects of $D'_{\mu\nu}$ in scattering experiments may be
similarly difficult. It is impossible to test for a purely timelike
$\kappa^{\mu}$ in the center of mass frame of
a two-body collision. The momenta $\vec{p}_{1}$ and $\vec{p}_{2}$
of the incoming particles must be equal and opposite; when the corresponding currents
are contracted with the $\epsilon$-tensor in the
effective photon propagator, they produce a vanishing result.

In summary, we have introduced a new interaction that couples right- and
left-cir\-cu\-lar\-ly polarized photons to moving charges differently. Although the
helicity of a single photon is invariant under rotations and Lorentz boosts, the
new interaction still violates Lorentz invariance, because of the way it interacts
with electrostatic effects. The interaction is also nonlocal, for a similar reason.
However, the propagator for the modified theory (which we have determined exactly) is
transverse and describes only two propagating photon modes per wave vector. The
propagator gives rise to such novel effects as forces between perpendicular currents.
The best bounds on $\kappa$, which parameterizes the strength of the
helicity-dependent coupling, come from precision polarimetry of synchrotron radiation
and are at the $2\times 10^{-3}$ level.

\end{document}